\documentclass[fleqn,usenatbib]{mnras}

\usepackage[T1]{fontenc}

\DeclareRobustCommand{\VAN}[3]{#2}
\let\VANthebibliography\thebibliography
\def\thebibliography{\DeclareRobustCommand{\VAN}[3]{##3}\VANthebibliography}

\usepackage{graphicx}	
\usepackage{amsmath}	
\usepackage{amssymb}	
\usepackage{newtxtext,newtxmath}

\title[Gravitational waves from neutron stars]{Gravitational waves from spinning neutron stars as not-quite-standard sirens}

\author[M. Sieniawska \& D. I. Jones]{
Magdalena Sieniawska,$^{1}$\thanks{E-mail: magdalena.sieniawska@uclouvain.be}
David Ian Jones,$^{2}$
\\
$^{1}$Centre for Cosmology, Particle Physics and Phenomenology (CP3), Universit\'{e} catholique de Louvain, Chemin du Cyclotron 2, B-1348 Louvain-la-Neuve, Belgium \\
$^{2}$Mathematical Sciences and STAG Research Centre, University of Southampton, Southampton SO17 1BJ, United Kingdom
}

\date{Accepted XXX. Received YYY; in original form ZZZ}

\pubyear{2021}

\begin{document}
\label{firstpage}
\pagerange{\pageref{firstpage}--\pageref{lastpage}}
\maketitle

\begin{abstract}
As is well known, gravitational wave detections of coalescing binaries are standard sirens, allowing a measurement of source distance by gravitational wave means alone.  In this paper we explore the analogue of this capability for continuous gravitational wave emission from individual spinning neutron stars, whose spin-down is driven purely by gravitational wave emission.  We show that in this case, the distance measurement is always degenerate with one other parameter, which can be taken to be the moment of inertia of the star.   We quantify the accuracy to which such degenerate measurements can be made.  We also discuss the practical application of this method to scenarios where one or other of distance or moment of inertia is constrained, breaking this degeneracy and allowing a measurement of the remaining parameter. We consider a broad range of possible, unknown parameters, as well as we present results for the aLIGO and Einstein Telescope sensitivities. Our results will be of use following the eventual detection of a neutron star spinning down through such gravitational wave emission.  
\end{abstract}

\begin{keywords}
stars: neutron -- gravitational waves -- stars: distances
\end{keywords}


\section{Introduction}
\label{sec:intro}
Gravitational-wave (GW) astronomy has been one of the fastest-growing fields in modern astrophysics. The first GW detection of the binary black-hole (BH) system GW150914 \citep{Abbott2016} in 2015 opened a new channel to test theories about the sources of gravitational waves, cosmology, astrophysical processes and gravitation itself. So far, tens of coalescencing binaries - double BHs or double neutron stars (NSs) - have been detected \citep{Abbott2019, Abbott2020a} by the LIGO \citep{Aasi2015} and Virgo \citep{Acernese2014} instruments.

GWs from coalescencing binary systems are `standard sirens' \citep{Schutz1986} - the GW analog of an astronomical standard candle - as determination of their luminosity distance depends only on measurable quantities like amplitude, frequency and frequency derivative of the signal. Additionally, double NS mergers have electromagnetic counterparts that can allow identification of the host galaxy. Such multi-messenger observations allow determination of cosmological parameters like e.g. Hubble constant. Such an analysis \citep{Abbott2017a} was performed for the first multi-messenger detection, the GW170817 event \citep{Abbott2017b, Abbott2017c, Abbott2017d}.   It is even possible to use gravitational wave observations of binary black coalescences to constrain the Hubble constant, despite the absence of a uniquely identified host galaxy \citep{LVK_19_H0_DES, LVK_21_H0}.

In addition to inspirals and mergers of compact objects, there are other classes of objects that can produce gravitational radiation.  This includes long-lasting and almost-monochromatic emission from  isolated, spinning NSs. Such continuous gravitational waves (CGW) might be due to the steady rigid rotation of a triaxial star, whose triaxiality or ``mountain'' is supported by elastic and/or magnetic strains.  Alternatively, the emission may be due to oscillations in a rotating star, with r-mode oscillations a prime candidate.  See  \citet{Andersson2011, Lasky2015, Riles2017, Sieniawska2019} for relevant reviews. As the GW detectors improve their sensitivity and data analysis methods are constantly upgraded, CGW signals are considered as serious candidates for future detections.

In this work we investigate the possibility of using CGW sources as standard sirens.  So far pulsar distances have been determined using electromagnetic observations, mainly via dispersion measurements \citep{Donner2020}.  For the coalescencing binary NS systems it is possible to determine their distance directly from the GW observations. However, for the CGW sources, we show that distance estimation is always degenerate with one additional unknown parameter.  We nevertheless analyse the accuracy to which such measurements can be made, and comment on the extent to which additional information can be used to break this degeneracy.  We consider  CGW signals produced by  mountains and by r-modes oscillations. We consider a broad range of the possible parameters, including ellipticity, r-mode amplitude and initial rotational frequency.  We give results for the aLIGO detector\footnote{\url{https://dcc.ligo.org/LIGO-P1200087-v42}} \citep{Abbott2020a} and third-generation, planned Einstein Telescope\footnote{\url{http://www.et-gw.eu/index.php/etsensitivities}} \citep{Sathyaprakash2012}.  We show how our formulae can be modified to take cosmological red-shift corrections into account.  

There exists another, completely different method, whereby CGWs can be used to infer source distances, via parallax, as described in \cite{seto_05}.  This method has the advantage of not suffering from any degeneracies, but only works for relatively nearby sources, with distances of a kiloparsec or less (see figure 1 of \citealt{seto_05}).

The paper is composed as follows: in section~\ref{sect:meth} we introduce general information about gravitational radiation theory, our signal model and CGW detectability. We also compare distance estimation for coalescing binaries with that for NSs with mountains or r-modes oscillations, as well as considering cosmological corrections. We estimate errors in the measurement of relevant  signal parameters, to help assess when our ideas can be meaningfully applied. At the end of this section we motivate our assumptions and the parameter space used in the simulations. Section~\ref{sect:res} contains results of our simulations. In section~\ref{sect:discussion} we present some discussion, while in section~\ref{sect:conc} we conclude our work.

\section{Methods}
\label{sect:meth}

\subsection{Gravitational radiation theory}
\label{sec:grav} 
According to the general theory of relativity \citep{Einstein1916, Einstein1918}, GWs are perturbations in the curvature of space-time, travelling at the speed of light. The lowest radiating multipole is the quadrupole, as the emission from the monopole is forbidden by mass conservation and emission from the dipole by momentum conservation. The general expression for the GW amplitude strain tensor $h_{ij}$ at position $r$ is:
\begin{equation}
    h_{ij} = \frac{2G}{c^4 r}\ddot{Q}^{TT}_{ij}\left(t - \frac{r}{c}\right),
    \label{eq:hij_general}
\end{equation}
where $Q^{TT}_{ij}$ is the mass-quadrupole moment in the transverse-traceless (TT) gauge, evaluated at the retarded time $(t - r/c)$, $c$ and $G$ are the speed of light and gravitational constant, respectively. 

For the CGW emission from a rigidly rotating  triaxial star, the amplitude of the signal given by equation~(\ref{eq:hij_general}) can be conveniently parameterised in terms of \citep{Ostriker1969, Melosh1969, Chau1970, Press1972, Zimmermann1978}:
\begin{equation}
    h_{0, \rm tr} = \frac{4G}{c^4}\frac{1}{d}I_3\epsilon\omega_{\rm rot}^2,
    \label{eq:h0tr}
\end{equation}
where $d$ is the distance to the source, $I_3$ is the moment of inertia along the spin axis, $\epsilon$ is the ellipticity that measures how different from spherical shape
the body is, defined as $\epsilon = (I_2 - I_1)/I_3$ (where $I_1$ and $I_2$ are moments of inertia along axes perpendicular to $I_3$) and $\omega_{\rm rot} = 2 \pi f_{\rm rot}$ is the rotational (angular) frequency. For the triaxial ellipsoid model $f_{\rm GW} = 2 f_{\rm rot}$, where $f_{\rm GW}$ is the CGW frequency.

Assuming that the spin-down is driven by GW emission alone, one can use the conservation of energy to derive an expression for the rotational frequency derivative:
\begin{equation}
    \dot{\omega}_{\rm rot}= - \frac{32G}{5c^5}\omega_{\rm rot}^5 \epsilon^2 I_3,
    \label{eq:spindown_tr}
\end{equation}
Additionally assuming that the ellipticity $\epsilon$ is constant in time, the equation can be integrated to give:
\begin{equation}
    \omega_{\rm rot} (t)=\frac{\omega_{0, \rm rot}}{\left( \frac{128}{5} \frac{G}{c^5}\epsilon^2I_3\omega_{0, \rm rot}^4t +1 \right)^{1/4}},
\label{eq:omega_tr}
\end{equation}
where $\omega_{0,{\rm rot}}\equiv\omega_{\rm rot}(t=0)$ is the rotational frequency at the beginning of observations.

Another mechanism for producing CGWs are r-mode oscillations.  These are a subset of the inertial waves, caused by the Coriolis force acting as restoring force \citep{Rossby1939}. In NSs the r-modes can be amplified by the Chandrasekhar-Friedman-Schutz instability \citep{Chandrasekhar1970, Friedman1975, fs_78b, andersson_98}. This instability is driven by GW back-reaction - it tends to amplify hydrodynamic waves in the fluid components, which propagate in the opposite direction to that of the NS rotation, producing GWs. According to \citet{Owen2010}, the CGW strain amplitude for the r-modes case can be expressed as:
\begin{equation}
h_{0,\rm rm} = \sqrt{\frac{8\pi}{5}}\frac{G}{c^5}(\alpha MR^3 \tilde{J})\frac{1}{d}\omega_{\rm mode}^3,
\label{eq:h0rm}
\end{equation}
where $M$ and $R$ are mass and radius of the star, respectively, and the angular frequency of the mode (and also of their CGW emission) is $\omega_{\rm mode} = 2\pi f_{\rm mode}$ with $f_{\rm mode}\approx 4 f_{\rm rot} /3$ for the mode of interest.   The amplitude of the mode is parameterised by $\alpha$, a dimensionless constant \citep{owen_etal_98}, while $\tilde{J}$ is another dimensionless parameter, defined as:
\begin{equation}
\tilde{J} = \frac{1}{MR^4}\int\displaylimits_0^R \hat{\rho} r^6 dr,
\end{equation}
where $\hat{\rho}$ represents the mass density and $r$ the radial coordinate.

Similarly as for the triaxial ellipsoid case, spindown can be derived from the conservation of energy, assuming no other energy losses:
\begin{equation}
    \dot{\omega}_{\rm rot} = - \frac{2^{18}\pi G}{3^8 5^2 c^7} (\alpha MR^3\tilde{J})^2 \frac{1}{I_3} \omega_{\rm rot}^7 .
    \label{eq:spidown_rm}
\end{equation}
Assuming constant mode amplitude $\alpha$, this can be integrated to give:
\begin{equation}
\omega_{\rm rot}(t) = \frac{\omega_{0, {\rm rot}}}{\left( 1 + \frac{2^{19}\pi G}{3^7 5^2 c^7} t\omega_{0, \rm rot}^6(\alpha MR^3 \tilde{J})^2 \frac{1}{I_3}\right) ^{1/6}}.
\label{eq:omegat,rm}
\end{equation}

The detectability of the GW signal is given in terms of the signal-to-noise ratio ($\rho$, SNR), as explained in \citet{Moore2015}:
\begin{equation}
\rho^2 = \int\displaylimits_{f_{\rm GW,beg}}^{f_{\rm GW,end}} \left(\frac{h_c(f)}{h_n(f)}\right)^2 d(\ln f),
\label{eq:snr}
\end{equation}
where $f_{\rm GW,beg}$ and $f_{\rm GW,end}$ are the GW frequencies of the signal at the beginning an end of observational time, respectively. $h_{\rm c}(f)$ is characteristic amplitude, defined as:
\begin{equation}
h_{\rm c}(f) = 2f\cdot \vert \tilde{h}(f) \vert,
\label{eq:hc}
\end{equation}
where $\tilde{h}(f)$ is a Fourier transform of the CGW signal \citep{Finn1993}. The above equation can be averaged over sky location and source orientation \citep{Jaranowski1998}, resulting in the averaged characteristic amplitude $\langle h_{\rm c} (f) \rangle = \frac{2}{5} h_{\rm c} (f)$.  The quantity $h_{\rm n}$ is  an effective noise of the detector given by:
\begin{equation}
h_{\rm n}(f) = \sqrt{f\cdot S_h(f)},
\end{equation}
where $S_h$ is the amplitude spectral density (a measure of the sensitivity of the detector). 
The integration in Eq.~\ref{eq:snr} is over the frequency of the signal, from the value at the beginning of observation time, to the value on the end of observations.  For parts of our parameter space, there is considerable variation in spin frequency (and amplitude), hence the need for this integration.

\subsection{Distance estimation}
The idea to use coalescencing binaries as `standard sirens' and determine their distance directly from GWs observations has been known for a long time \citep{Schutz1986, Markovic1993}. For such signal, the GW amplitude is given by (neglecting the dependence on sky position and source orientation):
\begin{equation}
    h_{0, \rm bin} = \frac{4 \pi^{2/3} G^{5/3}}{c^4}( f_{\rm GW} \mathcal{M})^{5/3} \frac{1}{f_{\rm GW}}\frac{1}{d},
    \label{eq:h0_inspiral}
\end{equation}
where $d$ is the distance, $f_{\rm GW}$ the GW frequency and $\mathcal{M}$ is the chirp mass - a function of the component masses $M_1$ , $M_2$:
\begin{equation}
\mathcal{M} = \frac{(M_1 M_2)^{3/5}}{(M_1+M_2)^{1/5}}.
\end{equation}
During the inspiraling phase, when two stars are sufficiently far apart, the post-Newtonian approximation can be applied, which is an expansion of general relativity when the velocity of the objects is small compared to the speed of light. For the merger phase numerical relativity has to be applied. The (measurable) frequency derivative (during the inspiral phase) is related to the chirp mass as:
\begin{equation}
    \dot{f}_{\rm GW} = \frac{96}{5}\pi^{8/3}\left(\frac{G\mathcal{M}}{c^3}\right)^{5/3}f_{\rm GW}^{11/3}.
\label{eq:dotf_bin}
\end{equation}
Equations~(\ref{eq:h0_inspiral}) and (\ref{eq:dotf_bin}) contain the measurable quantities $f_{\rm GW}$, $\dot{f}_{\rm GW}$ and $h_0$, and also the two (unknowns) $d$ and $\mathcal{M}$.  it follows one can solve for the two unknowns.  In particular, eliminating $\mathcal{M}$ gives:
\begin{equation}
    h_{0, \rm bin} = \frac{5c}{24\pi^2}\frac{1}{d} \frac{\dot{f}_{\rm GW}}{f_{\rm GW}^3},
\end{equation}
which makes it clear that the GW measurement of $h_{0, \rm bin}$, $\dot{f}_{\rm GW}$ and  $f_{\rm GW}$ allow calculation of the unknown $d$.  

Here we perform similar manipulations for CGW sources.  In the case of emission from a mountain,  equations~(\ref{eq:h0tr}) and (\ref{eq:spindown_tr}) give the signal amplitude and frequency evolution.  These two equations contain \emph{three} unknowns, $d, I_3$ and $\epsilon$.  This means that, unlike the binary case, we cannot solve for $d$.  The best we can do is to eliminate one of these three quantities, so that some combination of the other two remains.  As uncertainties in $I_3$ are smaller than for $\epsilon$ (as discussed later), we decided to eliminate $\epsilon$, leaving $d$ and $I_3$:
\begin{equation}
    h_{0, \rm tr} = \sqrt{\frac{5G}{2c^3}} \sqrt{\frac{\dot{\omega}_{\rm rot}}{\omega_{\rm rot}}} \frac{\sqrt{I_3}}{d}.
    \label{eq:h0tr_tau}
\end{equation}
We  see that we in fact can measure the combination $\sqrt{I_3}/d$, so the distance is degenerate with the moment of inertia.

One can perform similar manipulations for CGW emission from r-modes, by combining equations~(\ref{eq:h0rm}) and (\ref{eq:spidown_rm}) and eliminating the $(\alpha MR^3 \tilde{J})$ factor:
\begin{equation}
  h_{0, \rm rm} = \sqrt{\frac{45G}{8c^3}} \sqrt{\frac{\dot{\omega}_{\rm rot}}{\omega_{\rm rot}}} \frac{\sqrt{I_3}}{d}.
  \label{eq:h0rm_tau}  
\end{equation}

Note that is was very much a free choice in deciding to eliminate $\epsilon$.  In the case of mountains, if we had instead decided to eliminate $I_3$ between equation~(\ref{eq:h0tr}) and (\ref{eq:spindown_tr}) we would instead have
\begin{equation}
     h_{0, \rm tr} = \frac{5G}{2c^3} \frac{\dot{\omega}_{\rm rot}}{\omega_{\rm rot}} \frac{1}{\epsilon d} ,
    \label{eq:h0tr_tau_no_d}
\end{equation}
i.e. we would obtain a constraint on the product $\epsilon d$.  Similarly, equation~(\ref{eq:spindown_tr}) gives a constraint on the combination $\epsilon^2 I_3$.  Similar alternative choices were possible in the case of r-modes.

\subsection{Cosmological corrections}
\label{sect:cosmo_corr}
Previously we consider sources inside our Galaxy. Here we focus on CGW emitter at much larger, cosmological distances. In the case of such sources, redshift factors affect measurable parameters. This effect, for binary inspirals, modifies equation~(\ref{eq:h0_inspiral}) in the following way \citep{Schutz1986, Markovic1993}:
\begin{equation}
    h_{0,{\rm bin}} = 
    \frac{4 \pi^{2/3} G^{2/3}}{c^4}( f_{\rm d,GW} \mathcal{M}_{\rm d})^{5/3} 
    \frac{1}{f_{\rm d, GW}}\frac{1}{d_{l}},
    \label{eq:h0_inspiral_redshift}
\end{equation}
where $d_l$ is a luminosity distance and $f_{\rm d,GW}$ is a frequency in the detector frame, related to the frequency in the source frame $f_{\rm s, GW}$ via the redshift $z$ as $f_{s,GW} = f_{d,GW}(1+z)$. $\mathcal{M}_{\rm d}$ is a detector frame (i.e.\ redshifted) chirp mass, related to the (non-redshifted) chirp mass as $\mathcal{M}_{\rm d} = (1+z) \mathcal{M}$. Additionally, the frequency evolution of the signal is expressed as:
\begin{equation}
\dot{f}_{\rm d,GW} = \frac{96}{5}\pi^{8/3}\left( \frac{G}{c^3} \right)^{5/3} f_{\rm d,GW}^{11/3} \mathcal{M}_{\rm d}^{5/3},
\end{equation}
where the frequency derivative in the detector frame, $\dot{f}_{d,GW}$, is related to the frequency derivative in the source frame, $\dot{f}_{\rm s,GW}$, as $\dot{f}_{\rm d,GW} = \dot{f}_{s,GW}/(1+z)^2$. 

With the above equations one can deduce that it is not possible to determine separately chirp mass (in a source frame), distance and redshift - for these independent observation are needed, e.g. from the electromagnetic telescopes.

Analogously, for the emission from triaxial neutron stars, cosmological corrections modify equation~(\ref{eq:h0tr_tau}) in the following way:
\begin{equation}
h_{0, {\rm tr}} = 
\sqrt{\frac{5G}{2c^3}}\sqrt{\frac{\dot{f}_{\rm d,GW}}{f_{\rm d,GW}}}\frac{\sqrt{I_{3, \rm d}}}{d_{l}} ,
\label{eq:cosmo_h0tr}
\end{equation}
where we have introduced the \emph{detector frame moment of inertia}, $I_{3, \rm d}$, related to its source frame value by $I_{3, \rm d} = I_3(1+z)^3$.  In the above equation we have three measurable quantities: $h_{0, \rm tr}$, $f_{\rm d,GW}$, $\dot{f}_{\rm d,GW}$ and two unknown ones: $d_{\rm l}$ and $I_{3, \rm d}$. The factor of $(1+z)^3$ is readily understood, when one remembers that a moment of inertia is essentially a mass weighted quadratically with distance, with each factor of mass and length contributing one factor of $1+z$.

The effect of redshift for r-modes enters in the same way:
\begin{equation}
h_{0, \rm rm} = 
\sqrt{\frac{45G}{8c^3}}\sqrt{\frac{\dot{f}_{\rm d,GW}}{f_{\rm d,GW}}}\frac{\sqrt{I_{3, \rm d}}}{d_{l}}.
\label{eq:cosmo_h0rm}
\end{equation}

\subsection{Error estimation theory}
\label{sec:err} 
Previously we made some assumptions that all of our star's energy loss  goes into the CGW emission. Given this, one needs to attempt to test this using the gravitational wave observations alone.   The parameter that gives information about the mechanism behind the energy loss is the \emph{braking index}, defined as:
\begin{equation}
    n = \frac{\omega_{\rm rot} \ddot{\omega}_{\rm rot}}{\dot{\omega}_{\rm rot}^2}.
\end{equation}
For example, for the case of a mountain of constant size, $n=5$, while for r-modes of fixed amplitude $n=7$. 

To make quantitative estimates of how accurately we can measure the parameters of a signal, we will use the signal model presented in \citet{Jaranowski1999}.  This models the frequency evolution as a Taylor series, but neglects dependence of the signal on the source's sky location and on the orientation of its spin axis (see Discussion). In this model the phase of the signal is given as a polynomial, including terms up to the second frequency derivative:
\begin{equation}
    \Psi = \phi_0 
    + 2\pi \left[ f_{\rm GW}t + \frac{\dot{f}_{\rm GW}t^2}{2} + \frac{\ddot{f}_{\rm GW}t^3}{6}\right],
\end{equation}
where $t$ is an arbitrary time and $\phi_0$ is a initial phase, for simplicity set to be 0.

By using the  Fisher information matrix from \citet{Jaranowski1999}, the variance of the braking index estimation from the CGW detection is given by:
\begin{equation}
\begin{aligned}
\mathrm{var}(n) = &  \frac{1}{(\rho \pi)^2}\Bigg[ \frac{300}{T^2}\frac{\ddot{f}^2 _{\rm GW}}{\dot{f}^4_{\rm GW}} + \frac{25920}{T^4}\frac{f_{\rm GW}^2 \ddot{f}_{\rm GW}^2}{\dot{f}^6 _{\rm GW}} \\
& +\frac{25200}{T^6} \frac{f_{\rm GW}^2}{\dot{f}^4 _{\rm GW}} + \frac{5400}{T^3} \frac{f_{\rm GW}\ddot{f}^2_{\rm GW}}{\dot{f}^5_{\rm GW}} \\
&  +\frac{5040}{T^4} \frac{f_{\rm GW}\ddot{f}_{\rm GW}}{\dot{f}^4_{\rm GW}} +\frac{50400}{T^5} \frac{f_{\rm GW}^2\ddot{f}_{\rm GW}}{\dot{f}^5_{\rm GW}}  \Bigg],
\end{aligned}
\label{eq:var_n}
\end{equation}
where $T$ is the observation time.

Similarly, we can estimate the variance of the distance estimation for the CGW triggered by the mountain on the NS surface:
\begin{equation}
\begin{aligned}
\mathrm{var}\left(\frac{d}{\sqrt{I_3}}\right) = & \frac{5G}{4c^3}\frac{1}{(\rho \pi h_{0, \rm tr})^2}\biggl[ \frac{75\dot{f}_{\rm GW}}{T^2 f_{\rm GW}^3}+  \frac{1620}{T^4 f_{\rm GW} \dot{f}_{\rm GW}} \\
& +  \frac{\pi^2 \dot{f}_{\rm GW}}{f_{\rm GW}} + \frac{675}{T^3 f_{\rm GW}^2}  \biggr] 
\end{aligned}  
\label{eq:var_d_tr}
\end{equation}
and for the r-modes is:
\begin{equation}
\begin{aligned}
\mathrm{var}\left(\frac{d}{\sqrt{I_3}}\right)= & \frac{45G}{8c^3}\frac{1}{(\rho \pi h_{0,\rm rm})^2 }  \biggl[ \frac{75\dot{f}_{\rm GW}}{T^2 f_{\rm GW}^3}+  \frac{1620}{T^4 f_{\rm GW} \dot{f}_{\rm GW}} \\
& +  \frac{\pi^2 \dot{f}_{\rm GW}}{f_{\rm GW}} + \frac{675}{T^3 f_{\rm GW}^2}\biggr].
\end{aligned}
\label{eq:var_d_rm}
\end{equation}
In the section~\ref{sect:res} we present results in terms of standard deviations $\sigma$, which is a square root of the variances given in equations~(\ref{eq:var_n}), (\ref{eq:var_d_tr}) and (\ref{eq:var_d_rm}).

Note that some of the stars considered in the results below spin down significantly over the duration of the observation.  This Taylor series-based error analysis will not be accurate for such stars.  It will, however, be accurate for those stars with smaller ellipticities and lower birth frequencies that consequently spin down only a little.  Our analysis should therefore be robust at identifying the threshold between those stars whose emission and spin down is strong enough for our analysis to apply, and those where it is not.

\subsection{Assumptions and parameter space}
In the analysis presented in this work, we make the following assumptions:
\begin{enumerate}
    \item To consider a signal as detectable, we require a signal-to-noise ratio $\rho\geq20$. 
    Such a $\rho$ is currently quite optimistic for previously unknown sources.  However, in the future, when the ET will be operating, with corresponding increases in computational power available for searches, the detection of such signals will be more realistic.
    \item To confirm that the energy loss is transferred mostly to the CGW radiation, the braking index has to be close to $5$ (for the mountains) or to $7$ (for the r-modes). To enforce this, we require that braking index estimation error $\sigma(n)\leq0.5$.
    \item Here we assume that the ellipticity $\epsilon$ and the r-mode amplitude $\alpha$ are constant.
    \item We assume that observation time in our simulations is equal $1$ year, which is comparable with the previous LIGO and Virgo observing runs.
\end{enumerate}

Additionally, we will consider a broad range of possible birth spin frequencies. The maximum allowed spin frequency, above which the centrifugal forces causes mass shedding and destroy the star, is known as the Keplerian frequency. Its exact value is not known, as it depends on the equation of state of the NS, however it is reasonable to limit ourselves to $1500$ Hz; see e.g.\ \citet{Haensel2007}. For the mountain case, the  rotational frequency $f_{\rm rot}$ is related to the GW frequency $f_{\rm GW}$ as $f_{\rm GW}=2f_{\rm rot}$ and for  r-modes as $f_{\rm GW} =\frac{4}{3}f_{\rm rot}$, so we limit our simulations to (initial) GW frequencies of $3000$ Hz and $2000$ Hz, respectively. We do not consider the effect of proper motion of the source; this would be important only for very close/high velocity stars; see \citet{covas_21}.

We also consider a broad range of possible ellipticities, $\epsilon$. There have been several studies of the maximal ellipticity,  for multiple equations of states, for Newtonian and relativistic stars, see e.g.\ \citet{Ushomirsky2000, Owen2005, Haskell2007, Mannarelli2007, Knippel2009, GJS_12,  Johnson-McDaniel2013, GAJ_21, GA_21}. For a `typical' NS a maximum  ellipticity in the range $10^{-6} - 10^{-7}$ seems reasonable.  However, for more exotic states of matter (like superconducting quark matter), the maximum ellipticity can reach value of $10^{-1}$. For this reason, we consider ellipticities in the range between $10^{-1}$ up to $10^{-7}$ in this work.  Note that the above estimates are for \emph{maximum} ellipticities; the actual ellipticity depends upon the geological history and/or magnetic field configuration of the star.

Similarly, the value of the r-mode amplitude $\alpha$ is a not tightly constrained from theory; see \citet{arras_etal_03, brink_etal_04, bondarescu_etal_09}.   Here we investigated possible values of $\alpha$ from $10^{-1}$ to $10^{-6}$; see Discussion for further comment.

\section{Results}
\label{sect:res}
We simulated the expected signal-to-noise ratio $\rho$ for the wide range of possible triaxial stars and stars with r-modes.  For triaxial stars, we considered a range of ellipticities ($\epsilon = 10^{-1} - 10^{-7}$) and GW initial frequencies - frequencies at the beginning of observations ($f_{\rm GW,beg} = 100 - 3000$ Hz).
For r-modes we assumed some canonical parameters of the NS in the $(\alpha MR^3 \tilde{J})$ expression: according to \citet{Owen1998}, we put $\tilde{J}=1.635\cdot10^{-2}$, $M=1.4$ M$_{\odot}$, $R = 12.53$ km, and $I_3=10^{45}$ g$\cdot$cm$^2$. We allowed $\alpha$ to be in range between $10^{-1}$ to $10^{-6}$. For the r-modes we had $f_{\rm GW,beg}$ ranging from $70$ to $2000$ Hz. 

For both mountains and r-modes, we made a strong assumption that all energy loss is only due to the CGW emission, as well as assuming that the amplitude parameters - $\epsilon$ and $\alpha$ - are constant in time.  We considered a few different distances, all corresponding to sources inside our Galaxy ($d < 30$ kpc).

\begin{figure}
	\includegraphics[width=\columnwidth]{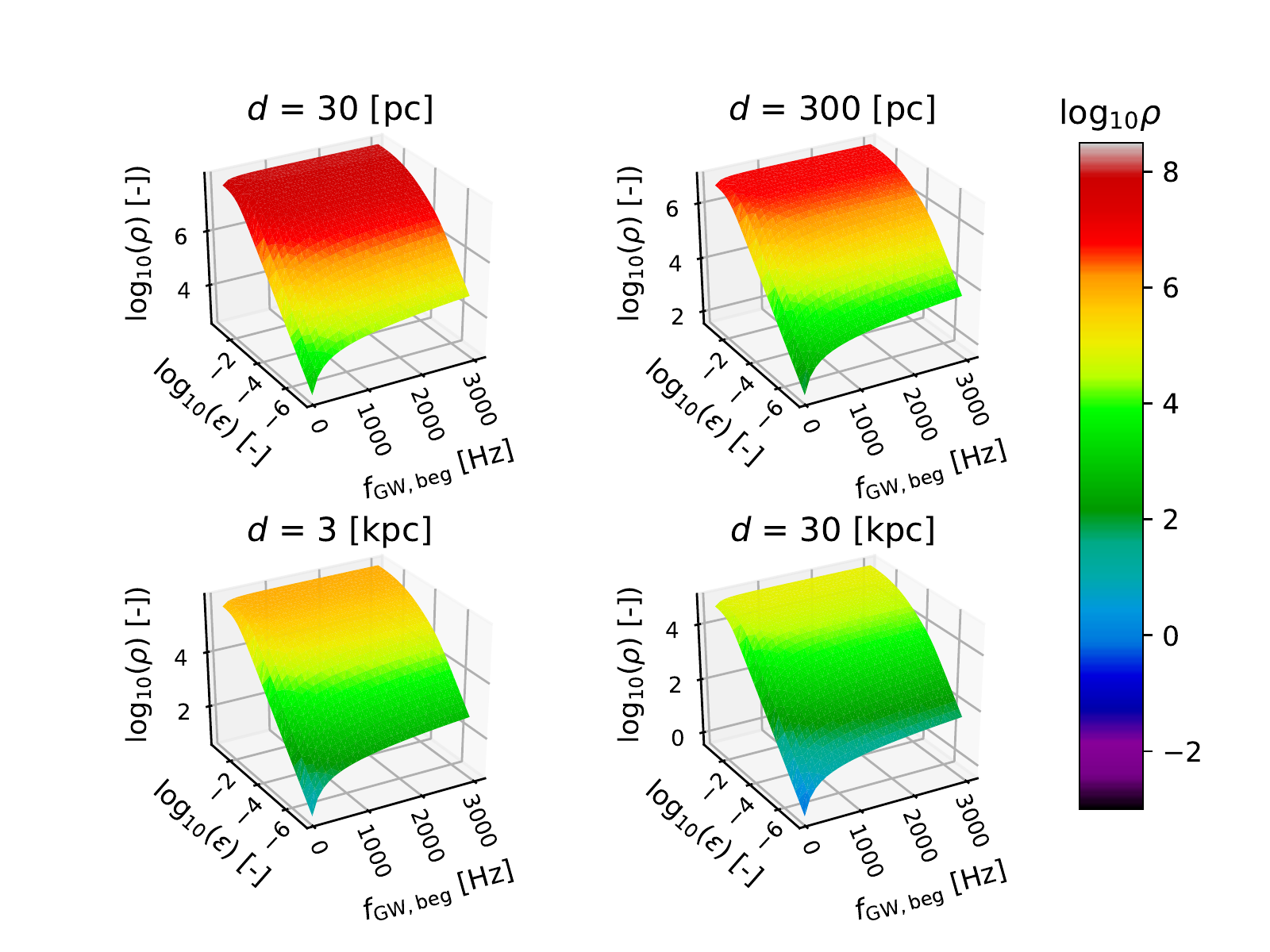}
    \caption{Expected signal-to-noise ratio $\rho$ (colourbar and z-axes) in the ET for the wide range of possible $\epsilon$ and $f_{\rm GW,beg}$, for the mountain case, at four different distances $d$. }
    \label{fig:3d_tr}
\end{figure}

\begin{figure}
	\includegraphics[width=\columnwidth]{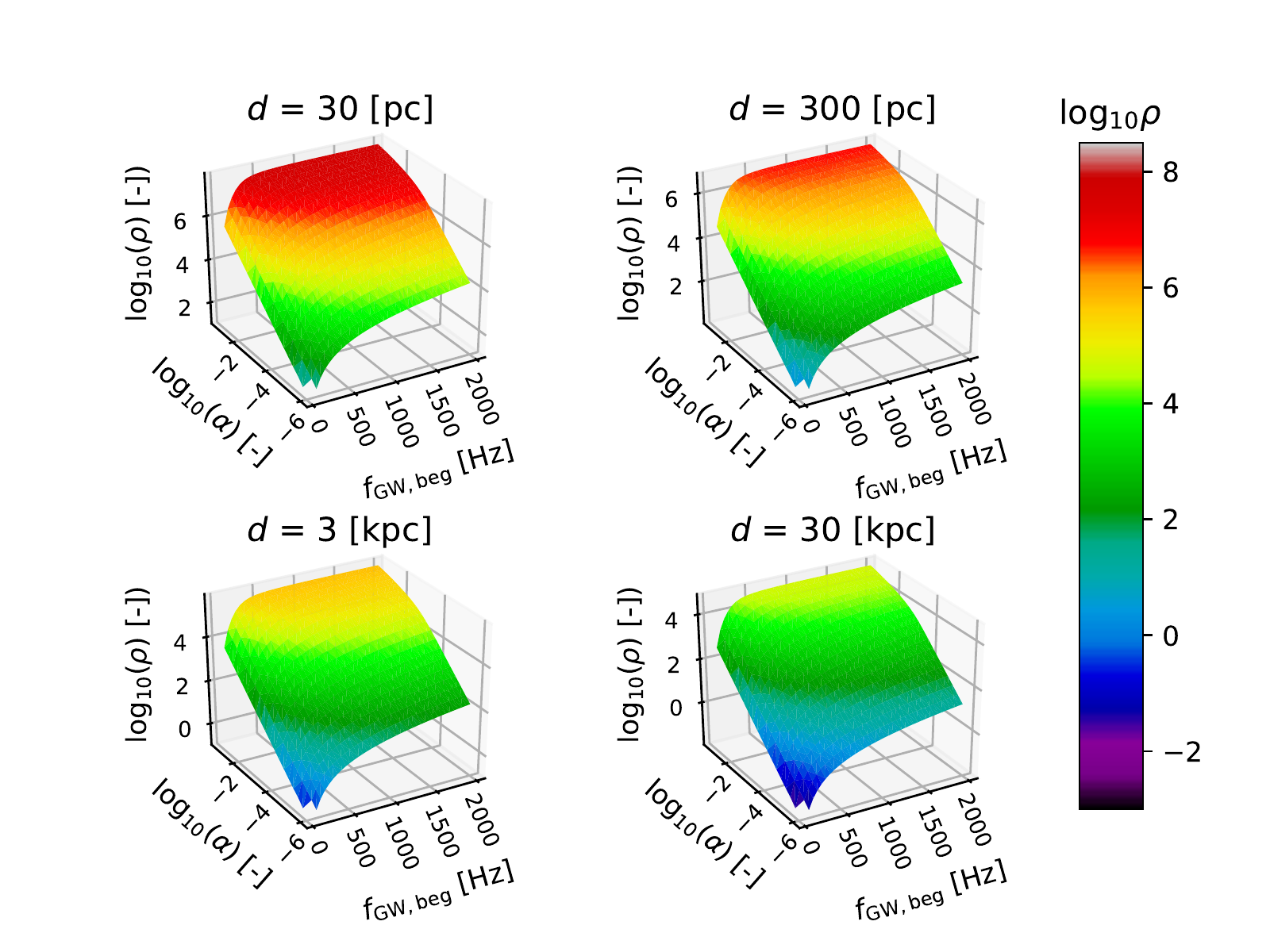}
    \caption{Same as in figure ~\ref{fig:3d_tr}, but for the r-modes case, for the wide range of possible $\alpha$ and $f_{\rm GW,beg}$. }
    \label{fig:3d_rm}
\end{figure}

The results presented in the figures in this section were produced for the ET sensitivity curves. However, we also performed simulations for the aLIGO sensitivity curve (design sensitivity) and these results are discussed later in this section. Signal-to-noise ratios for the broad ranges of parameters, for the mountains and r-modes cases, are shown in figures~\ref{fig:3d_tr} and \ref{fig:3d_rm}, respectively. For the large $\epsilon$ and $\alpha$ values, signals generate large $\rho$ for all considered distances and should be clearly visible in the GW detectors. Typically, results of the $\rho$ value for the aLIGO are about order of magnitude smaller in comparison with those for ET.  Note that the plots corresponding to the four different source distances in Figures~\ref{fig:3d_tr} and \ref{fig:3d_rm} are simple re-scalings of one another; we present them anyway for the reader's convenience, as the results given  for the same four star distances in all later plots are \emph{not} simple re-scalings of one another.

\begin{figure}
	\includegraphics[width=\columnwidth]{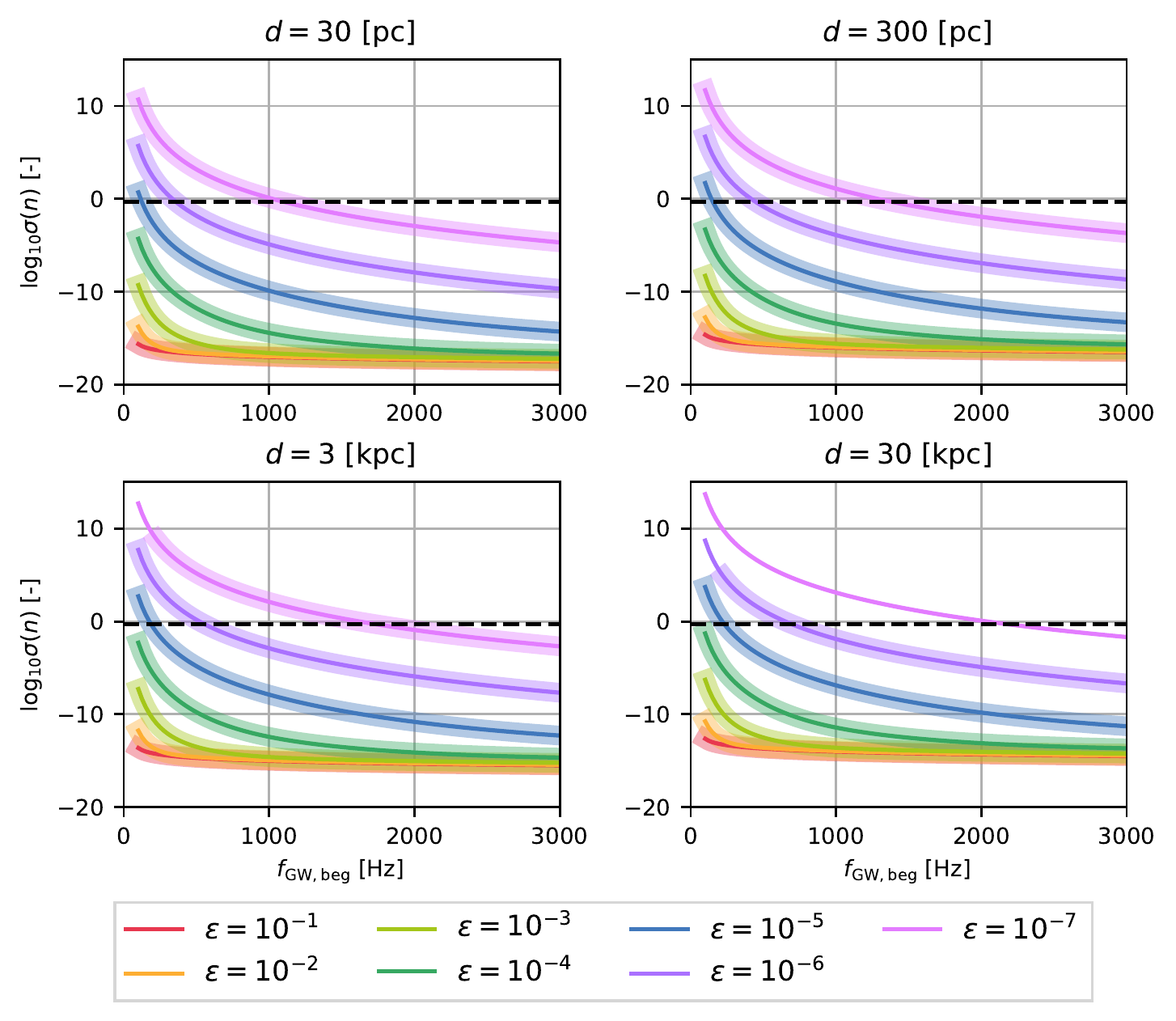}
    \caption{Estimation errors of the braking index $\sigma(n)$ for different Galactic distances $d$, initial GW frequencies $f_{\rm GW,beg}$ and ellipticities $\epsilon$, for the NSs with mountains. Thick transparent lines denote detectable signals, with $\rho>20$ (with the assumption that $100\%$ of the energy loss is transferred to the CGW emission). Black dashed horizontal line corresponds to the threshold $\sigma(n)=0.5$. }
    \label{fig:sdn_epsiloncuts}
\end{figure}

\begin{figure}
	\includegraphics[width=\columnwidth]{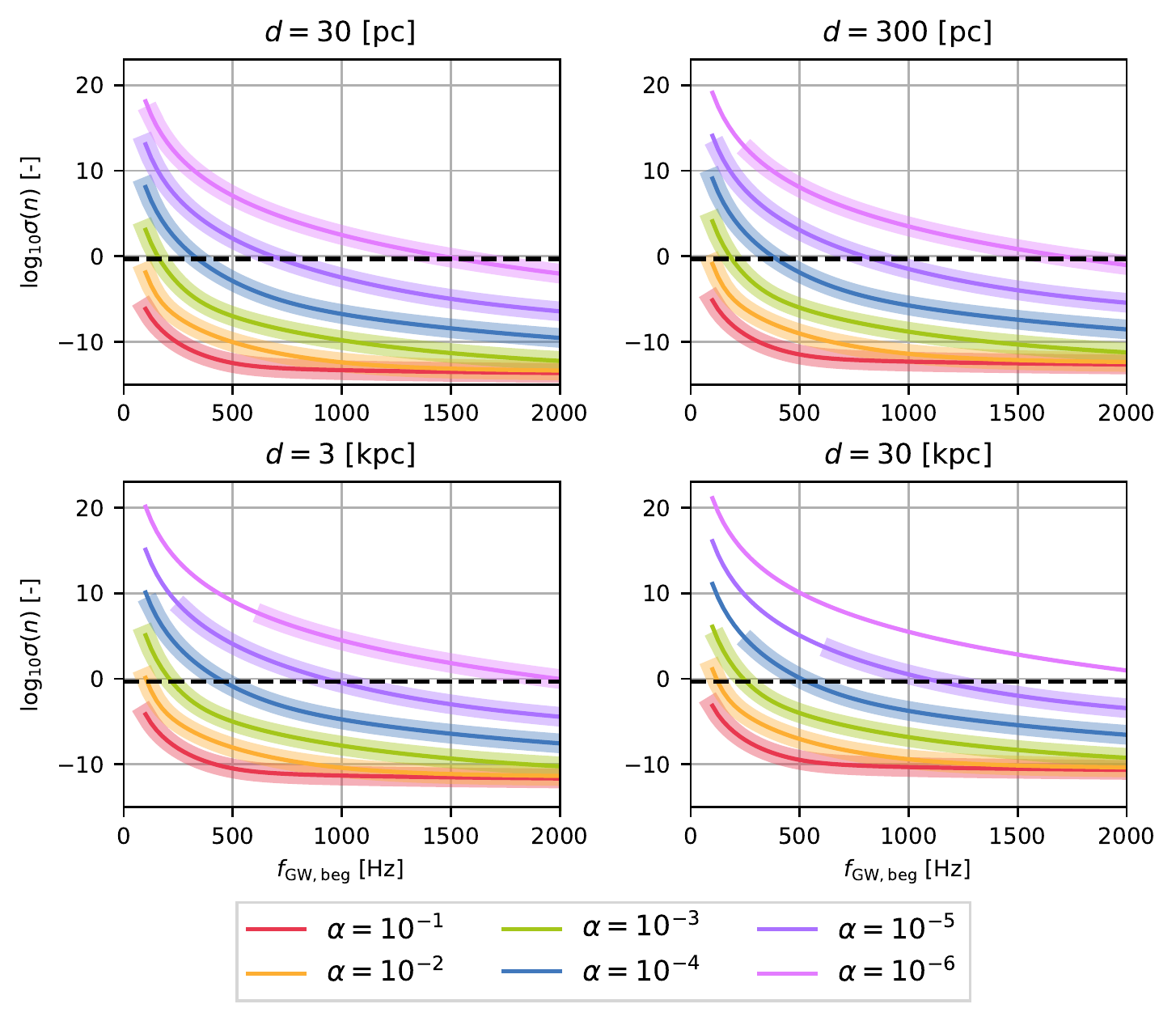}
    \caption{Same as figure~\ref{fig:sdn_epsiloncuts} but for the r-mode case. }
    \label{fig:sdn_alphacuts}
\end{figure}

We examined signals for which it is possible to estimate the braking index $n$ with satisfactory accuracy. We assumed that $\sigma(n)$ should be smaller or equal $0.5$ to confirm that the energy loss is due to the purely mountain emission ($n=5\pm 0.5$) or due to the r-modes ($n=7\pm 0.5$). Additionally, we require the signal to be detectable, with $\rho \ge 20$. Results for the mountains and r-modes cases are shown on figures~\ref{fig:sdn_epsiloncuts} and \ref{fig:sdn_alphacuts}, respectively. All NSs in our Galaxy with $\epsilon \geq 10^{-4}$, as well as with $\alpha \geq 10^{-1}$ fulfil our assumptions. Results for the aLIGO are very similar.

\begin{figure}
	\includegraphics[width=\columnwidth]{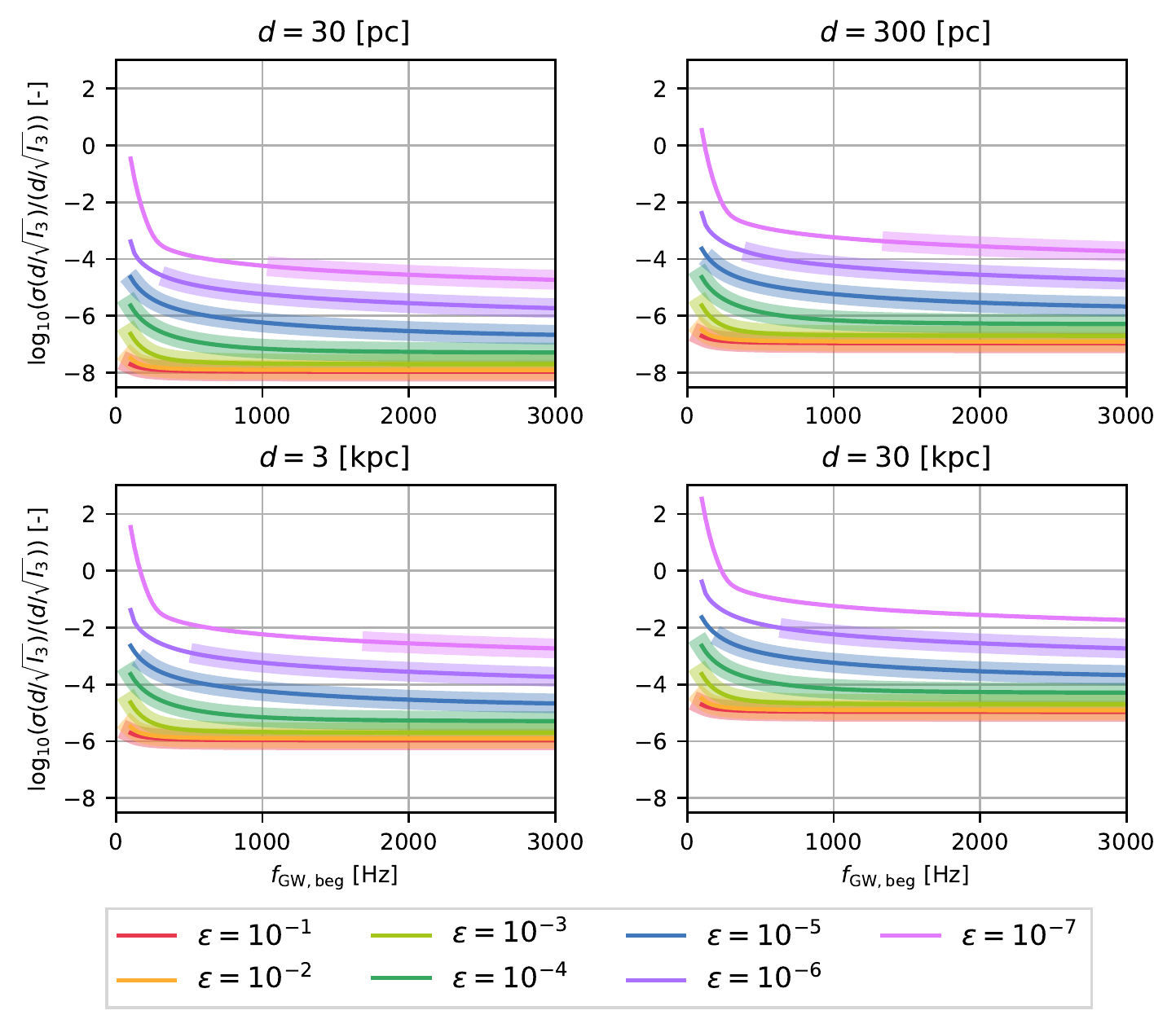}
    \caption{ Relative estimation errors of the quantity $\frac{\sigma(d/\sqrt{I_3})}{d/\sqrt{I_3}}$, in the case of CGW from a mountain. The thick transparent lines denote signals possible to detect, with $\rho>20$ and $\sigma(n)<0.5$. }
    \label{fig:sdd_epsiloncuts}
\end{figure}

\begin{figure}
	\includegraphics[width=\columnwidth]{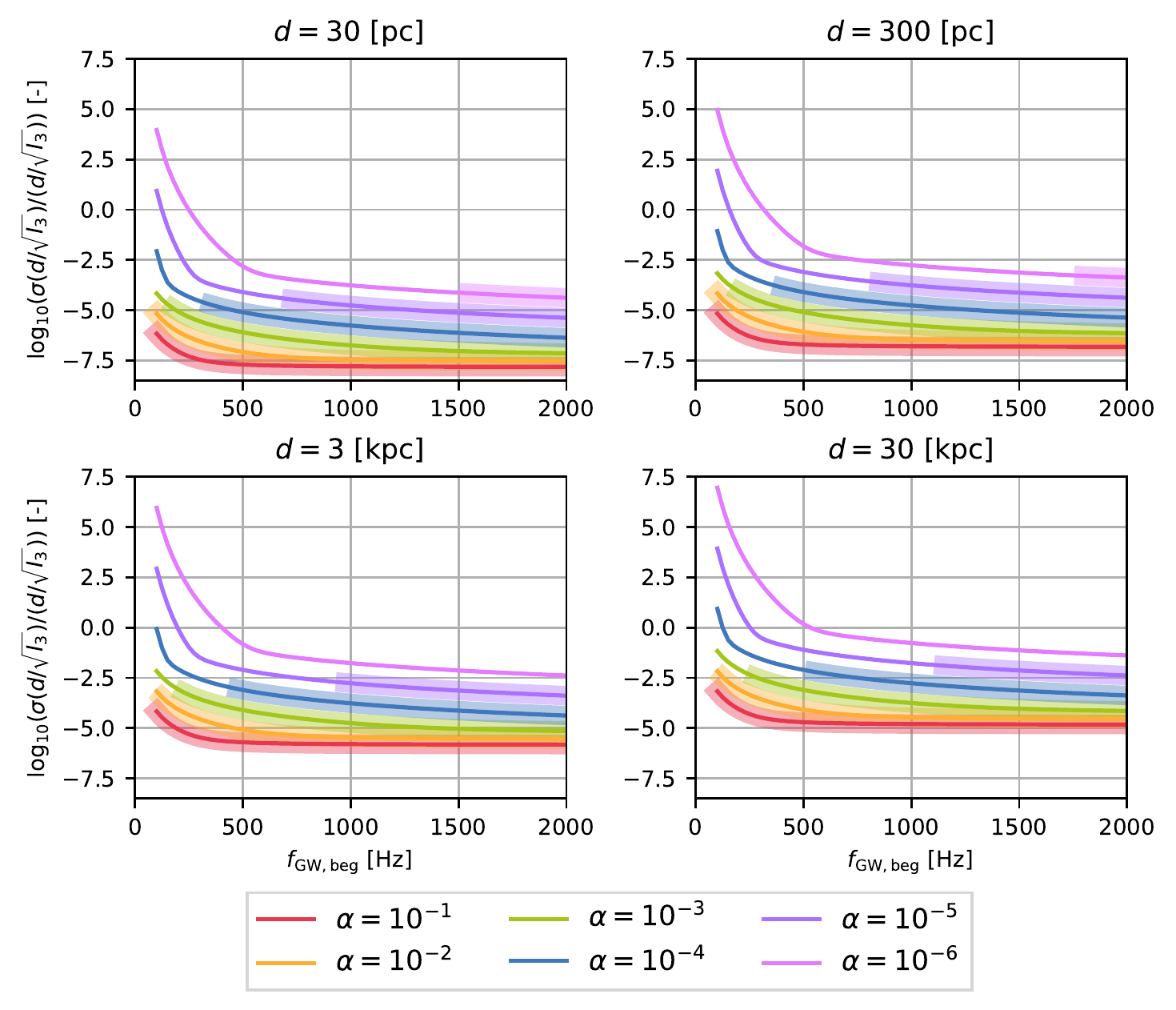}
    \caption{Same as for figure~\ref{fig:sdd_epsiloncuts} but for the r-mode case.}
    \label{fig:sdd_alphacuts}
\end{figure}

We simulated estimation errors of the degenerate combination  $d/\sqrt{I_3}$, for the signals that fulfil our assumptions ($\rho>20$ and $\sigma(n)<0.5$). Results for the mountains and r-modes cases are on figures~\ref{fig:sdd_epsiloncuts} and \ref{fig:sdd_alphacuts}, respectively. For all detectable signals ($\rho>20$ and $\sigma(n)<0.5$), in both cases, the relative error $\frac{\sigma(d/\sqrt{I_3})}{d/\sqrt{I_3}}$ is smaller than $1\%$. In the case of aLIGO this relative error increase to $10\%$. 


\begin{figure}
	\includegraphics[width=\columnwidth]{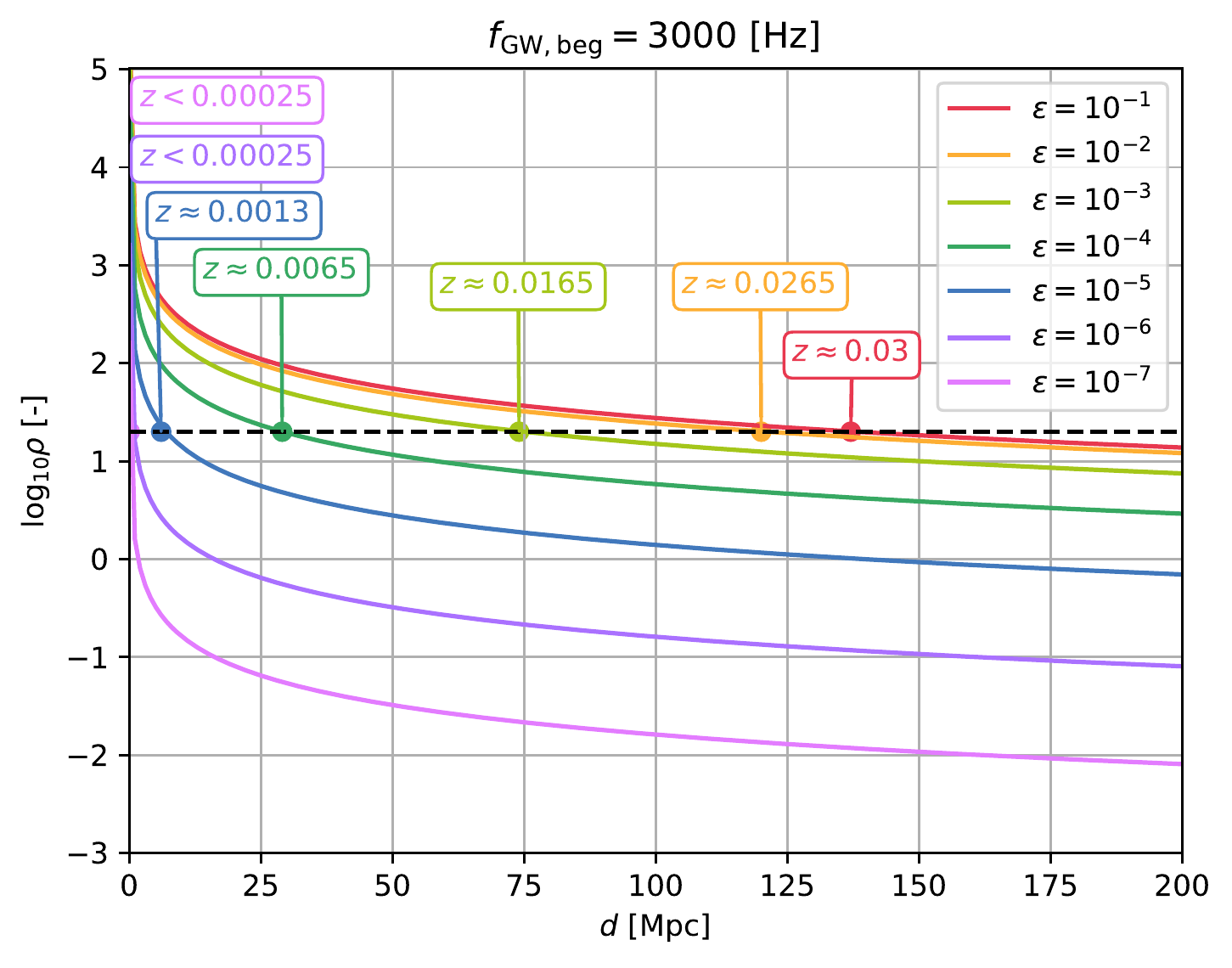}
    \caption{Expected signal-to-noise ratio $\rho$ as a function of distance $d$ for the NSs with $f_{\rm GW,beg}=3000$ Hz, for various $\epsilon$ values. The threshold $\rho=20$ is marked with the dashed line. Intersections between the dashed line and coloured ones correspond to the maximal detectable distance, for a given $\epsilon$. Corresponding redshifts are annotated on the plot.}
    \label{fig:rho_cosmo}
\end{figure}

We also investigated how inclusion of the redshift factor $z$ may influence the detectability of the signals, in the case of the sources at large, cosmological distances. By using equation~(\ref{eq:snr}) we calculated the signal-to-noise ratio for the NSs with the highest CGW amplitude within our frequency range ($f_{\rm GW,beg}=3000$ Hz for the mountains and $f_{\rm GW,beg} = 2000$ Hz for the r-modes), for various $\epsilon$ and $\alpha$. To consider the signal as a detectable one we put $\rho$ threshold to be 20. Results for the ET sensitivity curve and triaxial ellipsoid model are shown on figure~\ref{fig:rho_cosmo}. For the detectable sources, maximal corresponding distances $d_{\rm max}$ and redshifts\footnote{\url{https://www.kempner.net/cosmic.php}} (for the assumed $H_0=67.04$, $\Omega_m=0.3183$ and $\Omega_{\Lambda}=0.6817$) are shown in table~\ref{tab:redshift}. 

\begin{table}
\centering
\caption{Maximal distance $d_{\rm max}$ of the detectable CGW source for a given ellipticity $\epsilon$ or r-mode amplitude $\alpha$, corresponding redshift $z$ and its influence on the CGW amplitude measurement error.}
\begin{tabular}{|c||c|c|c|}
\hline
\hline
\multicolumn{4}{|c|}{Mountains} \\
\hline 
\hline
$\epsilon$ [-] & $d_{\rm max}$ [Mpc] & $z$ [-] & $h_{0, \rm tr}$ error\\
\hline
$10^{-1}$ & 137 & 0.03 & $\approx 9\%$ \\
$10^{-2}$ & 120 & 0.0265 & $\approx 8\%$ \\
$10^{-3}$ & 74 & 0.0165 & $\approx 5\%$ \\
$10^{-4}$ & 29 & 0.0065 &  $\approx 2\%$ \\
$10^{-5}$ & 6 & 0.0013 & $<1\%$ \\
$10^{-6}$ & $<1$ & $<0.00025$ & $<<1\%$ \\
$10^{-7}$ & $<1$ & $<0.00025$ & $<<1\%$ \\
\hline
\hline
\multicolumn{4}{|c|}{R-modes} \\
\hline 
\hline
$\alpha$ [-] & $d_{\rm max}$ [Mpc] & $z$ [-] & $h_{0, \rm rm}$ error\\
\hline
$10^{-1}$ & 86 & 0.019 & $\approx 6\%$ \\
$10^{-2}$ & 45 & 0.01 & $\approx 3\%$ \\
$10^{-3}$ & 18 & 0.004 & $\approx 1\%$ \\
$10^{-4}$ & 3 & 0.0007 &  $<1\%$ \\
$10^{-5}$ & $<1$ & $<0.00025$ & $<<1\%$ \\
$10^{-6}$ & $<1$ & $<0.00025$ & $<<1\%$ \\
\hline
\end{tabular}
\label{tab:redshift}
\end{table}

Also, in  table~\ref{tab:redshift} we show the percentage change in the CGW amplitude estimation due to the cosmological corrections. As $h_0$ is proportional to $\rho$, the same change will we visible in signal-to-noise ratio. Note that the distances to these sources are sufficiently small that it would be important to correct for peculiar velocities when attempting to measure $H_0$, as was the case for the binary neutron star detection GW170817 \citep{Abbott2017a}.   Clearly, the cosmological corrections are at most a few percent for the scenarios considered here, and so will not be very important for such detections. We will therefore not consider such corrections any further here.

\section{Discussion}
\label{sect:discussion}

We have shown that measurement of CGWs can allow estimation of combinations of distance, NS moment of inertia $I_3$ and ellipticity $\epsilon$ (or r-mode amplitude $\alpha$).  If one chooses to eliminate the highly uncertain $\epsilon$ (or $\alpha$), we are left with an estimate of $d / \sqrt{I_3}$.  This analysis is only applicable in situations were the spin-down is entirely (or almost entirely) driven by the CGW emission.  This means it would not make sense to apply our results to the (future) detection of GWs from a known young pulsar (e.g. the Crab, or Vela), where we already know from current CGW non-detections that only a fraction of the total spin-down energy can be radiated via the CGW channel \citep{LVK_08_beating_Crab, LVK_08_beating_Vela, Abbott2020c}.  The situation for known millisecond pulsars is less clear.  The fact that we see them as pulsars at all indicates that at least some of their spin-down energy is radiated electromagnetically, but the possibility of this being only a small fraction of the total energy budget cannot be discounted.  Indeed, \citet{woan_etal_18} recently advanced tentative evidence that the fastest spinning pulsars, located at the bottom left of the well known $P$--$\dot P$ diagram, may be mainly spin-down via CGWs due to a minimum ellipticity of the order of a few times $10^{-10}$.  However, given the impossibility of measuring braking indices for such slowly spinning-down objects (small $\dot P$), independent confirmation of the CGW-dominance of the energy budget would seem to be difficult to obtain, so direct gravitational wave detection may be the only avenue for progress \citep{Abbott2020c}.

Rather, our results are most likely to be useful for stars that are electromagnetically quiet, certainly with no significant magnetic dipole spin-down.  Such stars are often known as ``gravitars'' \citep{palomba_05}.  In the event of such a detection, the real interest of our results lies in the extent to which one can use additional information to constrain the individual values of $d$, $I_3$ and $\epsilon$, as (for instance) the combination $d / \sqrt{I_3}$ is of little interest in itself.  

A number of CGW searches have targeted small well-localised regions of the Galaxy.  These so-called \emph{directed searches} have targeted globular clusters \citep{LVC_17_globular}, the Galactic Centre \citep{LVC_13_galactic_centre,Piccinni2020}, and supernova remnants \citep{LVC_19_SNRs, Lindblom2020, Papa2020, Abbott2021}.  In the event of a detection from such a search, and assuming the signal is consistent with GW driven spin-down, one could then combine the estimated value of $d / \sqrt{I_3}$ with the independently estimated distance to the region in question to obtain a constraint on $I_3$ and $\epsilon$.  To give an idea of what might be possible, we note that the distance to the Galactic centre is known to within $\sim 5\%$ \citep{francis_anderson_14}.  The distance to a typical globular cluster is know to $\sim 6 \%$ \citep{chaboyer_08}, while distances to some supernova remnants are reported with errors of $\sim 5 \%$ or better \citep{ranasinghe_leahy_18}.  These uncertainties would then combine with the uncertainty in $d / \sqrt{I_3}$ from the CGW detection (figures~\ref{fig:sdd_epsiloncuts} and \ref{fig:sdd_alphacuts})     to give uncertainties in $I$ and $\epsilon$.  

A number of other CGW searches have performed so-called \emph{all-sky} analyses, searching over all possible search directions, for currently unknown NSs \citep{LVC_19_allsky, Abbott2021b, Abbott2021c, Steltner2021}.   In the event of a detection from such a search, and assuming the signal is consistent with GW-driven spin-down, it is less easy to make progress.  {\it A priori}, a galactic NS might lie anywhere from tens of parsecs to tens of kiloparsecs from Earth, while its moment of inertia probably lies somewhere in the relatively small range $1$--$2 \times 10^{45}$ g cm$^2$         \citep{worley_etal_08}.     This means one could at least constrain the highly uncertain distance and ellipticity (or r-mode amplitude) to within a factor of $\sqrt{2}$ or so.  Alternatively, and probably rather optimistically, targeted follow-up electromagnetic observations could conceivably yield more information.  If the star is young and therefore hot, blackbody radiation might be detected.  Such electromagnetic emission would play no role in the spin-down energy budget, but, if one assumed blackbody emission over the entire stellar surface, gives an estimate of $d / R$ \citep{ozel_freire_16}.  The estimates of $d/ \sqrt{I_3}$ and $d / R$ could then be combined to give an estimate of $R / I_3$, which can then be use to constrain the equation of state.

Our method will also be of use if CGWs were detected from the post-merger remnant left behind after the merger of two NSs \citep{LVK_19_long_duration, sl_21_post-merger}, assuming that the post-merger CGW emission is consistent with GW-driven spin-down.  If the inspiral/merger itself were detected via GWs, a distance estimate would already be available from the inspiral/merger phase \citep{Schutz1986, Abbott2017a}, allowing us to break the degeneracy between $d$ and $I_3$, i.e.\ allow estimation of $I_3$ and $\epsilon$ of the post-merger remnant.  The inspiral/merger phase would also give information on the stellar masses of the two pre-merger stars \citep{Abbott2017b}, and therefore give an upper limit on the mass of the NS remnant.   The combination of $I_3$ and an upper limit on mass $M$ would then provide a constraint on the equation of state.  

We have made some very strong assumptions in our analysis, which we will now briefly critique.  We have of course assumed entirely CGW-driven spin-down.  This is a strong assumption, but one that might be realised if at least some NSs are born with no significant external magnetic fields, or have their magnetic fields buried during a prolonged phase of accretion.  In such cases there would be no electromagnetically-driven spin-down.

We have also used a highly simplified signal model (see section \ref{sec:err}) that neglected the dependence of the signal on the source's sky location and on the orientation of its spin axis.  Sky location can be expected to be measured extremely accurately in CGW, as errors in sky location affect the phase of the received signal.  However, the absence of the inclination angle $\iota$ from our signal model is probably more serious.  In the case of compact binary coalescence there is a significant degeneracy between distance measurement and $\iota$, and was a significant factor in determining the accuracy of the measurement of the Hubble constant following the binary NS coalescence GW170817 \citep{LVC_17_Hubble}.  Inclusion of this inclination angle would presumably increase the errors in our analysis, although the increase will be small for the strongest signals we consider.    

We have also assumed that the asymmetry in the star, responsible for producing the CGWs, does not change in time, i.e. constant $\epsilon$ for mountains, and constant $\alpha$ for r-modes.  Given that some of our stars spin-down significantly over the course of a typical observation period ($\sim 1$ year), this is questionable.  Indeed, glitch activity in known pulsars is known to correlate with spin-down rate \citep{espinoza_etal_11}.  In terms of the stellar structure, for mountains, the ellipticity $\epsilon$ is likely to be sustained either by strains in an elastic crust (or core), or by strong internal magnetic fields (see e.g. \citet{jones_02}).  The solid crust may well crack in response to the decreasing centrifugal deformation \citep{baym_pines_71}.  For CGWs from elastic mountains, this would lead to a more complex waveform, with separate sections having different amplitudes and spin-down parameters \citep{ashton_etal_17}.  This would complicate the detection \citep{ashton_etal_18}, but as long as the ellipticity were constant  within each inter-glitch section then our method can still be applied.  Mountains sustained by  internal magnetic field may be more robust to spin-down, although if the core neutrons are superfluid and core protons superconducting, the outward migration of neutron vortices that the spin-down requires may cause the superconducting flux tubes to be dragged outwards \citep{ruderman_etal_98}.  This would alter the internal magnetic field structure, which may in turn alter the ellipticity.  Such a process could manifest itself as some departure from the canonical $n=5$ braking index, and would complicate application of our method.

The assumption of a constant $\alpha$ for r-modes is also crude.   For our spinning down stars, if the r-mode is active at all, it is probably because it is subject to the Chandrasekhar-Friedman-Schutz (CFS) instability \citep{andersson_98, lindblom_etal_98}. In such a case, the mode would have grown it is so-called saturation amplitude $\alpha_{\rm s}$ \citep{Owen1998}.   While early models assumed a constant value for $\alpha_{\rm s}$ \citep{Owen1998}, more detailed modelling indicate more complex behaviour, with different analyses coming to somewhat different conclusions \citep{arras_etal_03, brink_etal_04, bondarescu_etal_09}.  If $\alpha_{\rm s}$ were to change only on timescales long compared to the spin-down timescale, our method can be employed straightforwardly.  If instead $\alpha_{\rm s}$ were to vary more rapidly, but a simple model of how it varies were to be used, perhaps of the form $\alpha_{\rm s} = \alpha_{\rm s}(f_{\rm rot})$,  one could still use our method, simply amending the formulae for $h_0$ and the energy conservation calculation for $f_{\rm rot}(t)$ to allow for non-constant $\alpha$.  

Clearly, out assumption of constant $\epsilon$ or $\alpha$ is a strong one, but is at least testable via a measurement of braking index.

\section{Conclusions}
\label{sect:conc}

We have shown that CGW emission from a spinning NS can be used to make measurements of combinations of the star's distance $d$, moment on inertia $I_3$ and ellipticity $\epsilon$ (or mode amplitude $\alpha$, in the case of r-modes).    This requires that the star be spinning down entirely through GW emission, and that the size of the deformation producing the GWs remains constant throughout the observation, i.e. constant ellipticity for mountains, constant mode amplitude for r-modes.  Our criteria for when our analysis can be applied to a real detection is that the signal is both detectable and that the braking index $n$ can be measured sufficiently accurately.  This last criterion is necessary so that one could have confidence that the evolution in spin frequency is indeed driven by CGW emission from a constant size deformation.

Given these assumptions and constraints we gave estimates of the accuracy to which the combination $d / \sqrt{I_3}$ could be measured by GW means alone. As shown in figures~\ref{fig:sdd_epsiloncuts} and \ref{fig:sdd_alphacuts}, the achievable accuracies are a function of distance, birth frequency, and (very strongly) of the size of the deformation.  To give a few plausible examples, for a star at a distance of $3$ kpc, with birth spin frequency $500$ Hz, and ellipticity $10^{-6}$, a fractional accuracy of $10^{-3}$ is possible.  For a similar star emitting GWs via r-modes  with $\alpha = 10^{-4}$, the corresponding fractional accuracy is at a similar level.

In the absence of any further information, the factor of two or so uncertainty in $I_3$ allows such measurements to be translated into factor of two or-so uncertainties in distance and ellipticity.  If further information is available, further conclusions could be drawn, e.g. an electromagnetically derived distance estimate would allow measurement of moment of inertia and ellipticity themselves.

Clearly, while CGW emission from spinning NS does not provide the clean strand siren measurements of compact binary coalescence, one can still extract some useful distance information from such GW detections.

\section*{Acknowledgements}

The authors would like to acknowledge useful exchanges with Greg Ashton and David Keitel.  DIJ acknowledges financial support from the Science and Technology Facilities Council (STFC, UK) via grant no. ST/R00045X/1.

\section*{Data Availability}
No new data were generated or analysed in support of this research.



\bibliographystyle{mnras}
\bibliography{biblio} 







\bsp	
\label{lastpage}
\end{document}